\documentclass{article}
\usepackage{amsmath, amssymb, amsfonts}
\title{Rindler horizon entropy from nonstationarity} 
\author{Hristu Culetu, \\Ovidius University, Dept.of Physics, \\B-dul Mamaia 124, 900527 Constanta, Romania, \\e-mail : hculetu@yahoo.com}

\begin{document}
\numberwithin{equation}{section}
\pagenumbering{arabic}
\maketitle
\newcommand{\fv}{\boldsymbol{f}}
\newcommand{\tv}{\boldsymbol{t}}
\newcommand{\gv}{\boldsymbol{g}}
\newcommand{\OV}{\boldsymbol{O}}
\newcommand{\wv}{\boldsymbol{w}}
\newcommand{\WV}{\boldsymbol{W}}
\newcommand{\NV}{\boldsymbol{N}}
\newcommand{\hv}{\boldsymbol{h}}
\newcommand{\yv}{\boldsymbol{y}}
\newcommand{\RE}{\textrm{Re}}
\newcommand{\IM}{\textrm{Im}}
\newcommand{\rot}{\textrm{rot}}
\newcommand{\dv}{\boldsymbol{d}}
\newcommand{\grad}{\textrm{grad}}
\newcommand{\Tr}{\textrm{Tr}}
\newcommand{\ua}{\uparrow}
\newcommand{\da}{\downarrow}
\newcommand{\ct}{\textrm{const}}
\newcommand{\xv}{\boldsymbol{x}}
\newcommand{\mv}{\boldsymbol{m}}
\newcommand{\rv}{\boldsymbol{r}}
\newcommand{\kv}{\boldsymbol{k}}
\newcommand{\VE}{\boldsymbol{V}}
\newcommand{\sv}{\boldsymbol{s}}
\newcommand{\RV}{\boldsymbol{R}}
\newcommand{\pv}{\boldsymbol{p}}
\newcommand{\PV}{\boldsymbol{P}}
\newcommand{\EV}{\boldsymbol{E}}
\newcommand{\DV}{\boldsymbol{D}}
\newcommand{\BV}{\boldsymbol{B}}
\newcommand{\HV}{\boldsymbol{H}}
\newcommand{\MV}{\boldsymbol{M}}
\newcommand{\be}{\begin{equation}}
\newcommand{\ee}{\end{equation}}
\newcommand{\ba}{\begin{eqnarray}}
\newcommand{\ea}{\end{eqnarray}}
\newcommand{\bq}{\begin{eqnarray*}}
\newcommand{\eq}{\end{eqnarray*}}
\newcommand{\pa}{\partial}
\newcommand{\f}{\frac}
\newcommand{\FV}{\boldsymbol{F}}
\newcommand{\ve}{\boldsymbol{v}}
\newcommand{\AV}{\boldsymbol{A}}
\newcommand{\jv}{\boldsymbol{j}}
\newcommand{\LV}{\boldsymbol{L}}
\newcommand{\SV}{\boldsymbol{S}}
\newcommand{\av}{\boldsymbol{a}}
\newcommand{\qv}{\boldsymbol{q}}
\newcommand{\QV}{\boldsymbol{Q}}
\newcommand{\ev}{\boldsymbol{e}}
\newcommand{\uv}{\boldsymbol{u}}
\newcommand{\KV}{\boldsymbol{K}}
\newcommand{\ro}{\boldsymbol{\rho}}
\newcommand{\si}{\boldsymbol{\sigma}}
\newcommand{\thv}{\boldsymbol{\theta}}
\newcommand{\bv}{\boldsymbol{b}}
\newcommand{\JV}{\boldsymbol{J}}
\newcommand{\nv}{\boldsymbol{n}}
\newcommand{\lv}{\boldsymbol{l}}
\newcommand{\om}{\boldsymbol{\omega}}
\newcommand{\Om}{\boldsymbol{\Omega}}
\newcommand{\Piv}{\boldsymbol{\Pi}}
\newcommand{\UV}{\boldsymbol{U}}
\newcommand{\iv}{\boldsymbol{i}}
\newcommand{\nuv}{\boldsymbol{\nu}}
\newcommand{\muv}{\boldsymbol{\mu}}
\newcommand{\lm}{\boldsymbol{\lambda}}
\newcommand{\Lm}{\boldsymbol{\Lambda}}
\newcommand{\opsi}{\overline{\psi}}
\renewcommand{\tan}{\textrm{tg}}
\renewcommand{\cot}{\textrm{ctg}}
\renewcommand{\sinh}{\textrm{sh}}
\renewcommand{\cosh}{\textrm{ch}}
\renewcommand{\tanh}{\textrm{th}}
\renewcommand{\coth}{\textrm{cth}}

\begin{abstract}
 Finite entropy and energy are obtained for the horizon of a Rindler observer on the grounds of the nonstatic character of the geometry beyond the horizon. Edery - Constantineau prescription is used to find the dynamical phase space of this particular spacetime. The number of microstates rooted from the ignorance of a Rindler observer of the parameter $t$ from the nonstationary region are calculated. The entropy expression is also obtained from the electric field on the Rindler horizon generated in the comoving system of a uniformly accelerated charge.
 
 We suggest that the gravitational energy density constructed by means of the horizon energy and using the Holographic Principle is proportional to $g^{2}$, similar with a result recently obtained by Padmanabhan.
\end{abstract}

\section{Introduction}
 The black hole (BH) physics after Bekenstein and Hawking has implied that there is a deep connection among gravitation, thermodynamics and quantum information theory. The Hawking formula for the BH horizon temperature is an evidence (it includes all the fundamental constants of physics).
 
 Since horizons block informations to certain observers, it seems reasonable to associate an entropy with any event horizon \cite{TP1}. If a family of observers have no access to a part of spacetime, then they will attribute an entropy to the gravitational field because of the degrees of freedom (DOF) which are hidden behind the horizon. Padmanabhan goes further and shows \cite{TP2} that the spacetime has microscopic DOF and the Einstein field equations in the continuum limit are to be obtained as the coarse-grained, thermodynamic limit of the (unknown) microscopic laws. Therefore, there should exist a relation similar to the equipartition law $E = (1/2) nk_{B}T$ connecting the spacetime energy, temperature and the number of the microscopic DOF within that spacetime.
 
 It is generally accepted that the BH horizon has entropy but there is no consensus whether this is valid for Rindler's or deSitter's horizon, too (the fact that the spacetime has a microstructure allows one to obtain the dynamics extremizing a suitable thermodynamic potential, for example entropy \cite{TP2}).
 
 Edery and Constantineau \cite{EC} showed that non-extremal BHs contain a nonstationary region hidden behind the event horizon where the Killing vector becomes spacelike. In their view, the Schwarzschild BH stems from the nonstatic interior region: it is a measure of an outside observer's ignorance of the value of the time $t\in(0, 2m)$ which labels a continuous set of classical microstates. The authors of [3] applied the above idea to Schwarzschild, Reissner - Nordstrom and Kerr BHs, stressing that the extremal BHs have zero entropy because they do not contain nonstationary regions (that corresponds to a single metric configuration). 
 
 We apply Edery-Constantineau ideas to the spacetime felt by a uniformly accelerated observer who possesses a horizon and, beyond it, the geometry is nonstationary. We further show that the number of DOF on the Rindler horizon is proportional to $1/g^{2}$ where $g$ is the observer constant acceleration. Moreover, the horizon has energy and entropy thanks to the nonstatic character of the metric beyond it. A timelike congruence of geodesic observers is endowed with expansion and shear, depending on the acceleration $g$. Throughout the paper we take $G = c = \hbar = k_{B} = 1$.\\

 \section{The nonstatic Rindler metric}
 In Ref. [4] we have shown how the well known Rindler geometry 
 \begin{equation}
 ds^{2} = -(1-gX)^{2} dT^{2} + dX^{2} + dY^{2} + dZ^{2}
 \label{2.1}
 \end{equation}
 is obtained from the usual Minkowski metric by the coordinate transformation 
  \begin{equation}
  x_{M} = (\frac{1}{g} - X) cosh~gT,~~~~t_{M} = (\frac{1}{g} - X)sinh~gT,~~~ |x_{M}| > |t_{M}| .
 \label{2.2}
 \end{equation}
 (the $X = const.$ observers move along the hyperbola $x_{M}^{2} - t_{M}^{2} = (1/g - X)^{2}$ where $(x_{M}, t_{M})$ are Minkowski coordinates.
 The transformation
  \begin{equation}
 1 - gX = \sqrt{1 - 2g \bar{x}} 
 \label{2.3}
 \end{equation}
 brings the geometry (2.1) under the form
  \begin{equation}
 ds^{2} = - (1 - 2g \bar{x}) d \bar{t}^{2} + \frac{d \bar{x}^{2}}{1 - 2g \bar{x}} + d \bar{y}^{2} + d \bar{z}^{2} 
 \label{2.4}
 \end{equation}
 where $X \prec 1/g,~ \bar{x} \prec 1/2g$ and $ \bar{t} \equiv T,~\bar{y} = Y,~\bar{z} = Z$. In the region $ \bar{x} > 1/2g$, $1-2g \bar{x}$ becomes negative, $\bar{x}$ - timelike and $\bar{t}$ - spacelike. Therefore, we replace $\bar{x}$ with $t$ and $\bar{t}$ with $x$. One obtains
  \begin{equation}
 ds^{2} = - \frac{dt^{2}}{2gt - 1} + (2gt -1) dx^{2} + dy^{2} + dz^{2} 
 \label{2.5} 
 \end{equation}
 where $t\succ 1/2g,~y = \bar{y}, ~z = \bar{z}$. The geometry (2.5) is flat, nonstationary and is valid beyond the Rindler horizon $X = 1/g$ (or $\bar{x} = 1/2g$). 
 
 It is worth noting that, for $\bar{x} > 1/2g$, $1 - gX$ from (2.1) becomes imaginary and, therefore, $(1 - gX)^{2} < 0$. Consequently, $x_{M}^{2} - t_{M}^{2} = (1/g - X)^{2} < 0$. In addition, $dX^{2} = - d\bar{x}^{2}/(2g \bar{x} - 1) < 0$ and (2.1) becomes time dependent ($\bar{x}$ is timelike) beyond $X = 1/g$ or $x_{M} = t_{M}$. In other words, the spacetime (2.5) is valid beyond the Minkowski causal horizon $x_{M} = t_{M}$. We indeed obtain, by means of the transformation \cite{HC1, AS} 
\begin{equation}
 \sqrt{2gt - 1} = g \tau,~~~~x = \tau, 
 \label{2.6} 
 \end{equation}
the metric
  \begin{equation}
 ds^{2} = -d \tau^{2} + g^{2} \tau^{2} d \eta^{2} + dy^{2} + dz^{2} 
 \label{2.7} 
 \end{equation}
which is nothing else but the nonstationary Milne geometry \cite{AS, DB} (or the degenerate Kasner geometry), well known from Cosmology and also from the Relativistic heavy Ions Colissions (RHIC), where $\eta$ represents the rapidity. We could, of course, have obtained it directly from (2.1).
  
 A similar situation is encountered when the Schwarzschild horizon $r = 2m$ is crossed - the timelike Killing vector becomes spacelike and the geometry inside the BH is nonstationary \cite{HC2}. 
 
 According to Edery-Constantineau prescription, the phase space beyond the Rindler horizon where the metric (2.5) is valid does not correspond to a single microstate but to a continuous set of states parameterized by the time $t$, with $t\in(1/2g, \infty)$. Therefore, Rindler's horizon must have an entropy due to the inaccessibility  to have informations about internal configurations beyond the event horizon \cite{JB}.\\

 \section{Congruence of timelike geodesics}
  Let us take a family of spacelike hypersurfaces $\Sigma$ of constant $t$. The induced metric on $\Sigma$ is given by
  \begin{equation}
  h_{ab} = g_{ab} + u_{a} u_{b},
  \label{3.1}
  \end{equation}
 where $u^{a} = (\sqrt{2gt-1}, 0, 0, 0)$ is the velocity field of the congruence ($h_{ab} u^{b} = 0$) which is orthogonal to $\Sigma$. The indices $a,b$ run from 0 to 3.The dynamical phase space ($h_{ab}, P^{ab}$) is defined by \cite{EC} 
 \begin{equation}
 P^{ab} = \frac{\sqrt{h}}{16 \pi} (K^{ab} - K h^{ab}),
 \label{3.2}
 \end{equation}
 where $P^{ab}$ is the momentum conjugate to $h_{ab},~K_{ab} = \dot{h_{ab}}/2N$ is the extrinsic curvature of $\Sigma$, $h = det(h_{ab}$ and $N(t)$ is the lapse function, that is $N = 1/\sqrt{2gt-1}$. We have
 \begin{equation}
 h_{ab} = (0, 2gt-1, 1, 1), ~~K_{xx} = g \sqrt{2gt-1},~~K_{yy} = K_{zz} = 0,~~K = \frac{g}{\sqrt{2gt-1}}.
 \label{3.3}
 \end{equation}
 Eq. (3.2) yields the only nonzero components
 \begin{equation}
 P^{yy} = P^{zz} = - \frac{g}{16 \pi}.
 \label{3.4}
 \end{equation}
 The Hamiltonian constraint $ ^{3}R + K_{ab} K^{ab} - K^{2} = 0$ is obeyed nontrivially  in the spacetime (2.5) because $K_{ab} K^{ab} = K^{2} = g^{2}/(2gt-1) \neq 0$ ($^{3}R$, constructed with $h_{ab}$, is vanishing). We also note that $P^{ab}$ 's are constants and depend only on the acceleration $g$.
 
 For the other kinematical parameters of the congruence one obtains (by means of the software package Maple - GRTensor)\\
 - the scalar expansion\\
 \begin{equation}
 \Theta \equiv \nabla_{a} u^{a} = \frac{g}{\sqrt{2gt-1}},
 \label{3.5}
 \end{equation}
 namely $\Theta = K \equiv K^{a}_{a}$.\\
 - the shear tensor\\
  \begin{equation}
 \sigma_{ab} = \frac{1}{2}(h_{b}^{c} \nabla_{c} u_{a}+ h_{a}^{c} \nabla_{c} u_{b})-\frac{1}{3} \Theta h_{ab}
  \label{3.6}
 \end{equation}
 has the nonzero components
 \begin{equation}
 \sigma^{x}_{x} = \frac{2g}{3 \sqrt{2gt-1}},~~\sigma^{y}_{y} = \sigma^{z}_{z} = - \frac{g}{3 \sqrt{2gt-1}},
 \label{3.7}
 \end{equation}
 with $\sigma^{2} \equiv \sigma^{ab} \sigma_{ab} = 2g^{2}/3(2gt-1)$. The acceleration $a^{b} = u^{a} \nabla_{a} u^{b} = 0$, showing that the congruence is geodesic, that is the ''static'' observer with $u^{i} = 0 ~(i = 1, 2, 3)$ move along a geodesic (the situation resembles the BH interior case, where $r = const.$ observers are geodesic \cite{DLC} 
 
 Let us noting that all the kinematical parameters vanish when $g = 0$ or when $t \rightarrow \infty$. In addition, the time variation of the scalar expansion 
 \begin{equation}
 \dot{\Theta} \equiv u^{a} \nabla_{a} \Theta = - \frac{g^{2}}{2gt-1}
 \label{3.8}
 \end{equation}
 is negative, i. e. $\Theta(t)$  decreases (otherwise the Raychaudhuri equation
  \begin{equation}
\dot{\Theta} - \nabla_{b} a^{b}+ \sigma^{2}- \omega^{2}+ \frac{1}{3} \Theta^{2} = - R_{ab} u^{a} u^{b}
\label{3.9}
\end{equation}
will not be satisfied (we have here $a^{b} = 0$, the Ricci tensor $R_{ab} = 0$ and the vorticity tensor $\omega_{ab} = 0$, with $\omega^{2} \equiv \omega^{ab} \omega_{ab}$)).\\

 \section{Event horizon entropy}
 Using the Edery - Constantineau paradigm, the entropy is a measure of the Rindler observer ignorance on the value of the parameter $t$ which labels the nonstationarity of the metric fields beyond the horizon. However, their model does not give us a method to calculate the entropy or the gravitational energy, in general.
 
 Having known that the entropy of the Rindler spacetime should be nonzero, we have nothing else to do than to take its expression from \cite{HC3} 
 \begin{equation}
 S = \frac{\pi}{4g^{2}}
 \label{4.1}
 \end{equation}
 Keeping in mind that the Rindler horizon temperature is $T = g/2 \pi$, we immediately obtain
 \begin{equation}
 E = 2TS = \frac{1}{4g}
 \label{4.2}
 \end{equation}
 for the energy of the Rindler spacetime (see also \cite{KM}). We may of course get the expression of the entropy directly from the nonstatic geometry (2.5). The boundary of the spacetime corresponds to $t = 1/2g$ (that is, the horizon $\bar{x}_{max} = 1/2g$). As Edery and Constantineau have shown, most of the energy contribution comes from a thin slice in the interior region near the event horizon (for the Schwarzschild BH), which corresponds in our case to the initial time $t_{min}$. Taking therefore $4 \pi t_{min}^{2}$ as the area of the initial surface, one obtains
 \begin{equation}
 S = \frac{1}{4} 4 \pi t_{min}^{2} = \frac{\pi}{4g^{2}},
 \label{4.3}
 \end{equation}
 as in Eq. (4.1).
 
  We are now in a position to find the number $n$ of the DOF (or the number of internal configurations)
of (2.4) due to the time dependence of the metric (a set of classical microstates which express the ignorance of an observer located in the static region). 

From the equipartition rule we have $E = (1/2) n T$. Using (4.2) and the relation for the horizon temperature $T$, we obtain
\begin{equation}
n = \frac{\pi}{g^{2}}
\label{4.4}
\end{equation}
 When all fundamental constants are recovered, (4.4) becomes $n = (c^{7}/G \hbar) (\pi/g^{2})$. For example, the value $g = 10 ~m/s^{2}$ leads to $n \approx 10^{102}$ bits. In other words, the nonstationarity of Rindler spacetimes (2.1) or (2.3) beyond the horizon $\bar{x} = 1/2g$ leads to that enormous value of the number of microstates.
 
  Incidentally, when we try to write down the energy density $\bar{\rho}$ rooted from $E$, the following expression is reached
  \begin{equation}
  \bar{\rho} = \frac{E}{V} = \frac{1}{4g} \frac{3}{4 \pi} g^{3} = \frac{3g^{2}}{16 \pi},
  \label{4.5}
  \end{equation}
  where we have used the Holographic Principle taking $E$ to be located uniformly on a sphere of radius $1/g$ (we stress here the special role played by the distance $1/g$ \cite{JWL}), instead of being concentrated on the horizon. But the relation (4.5) for the energy density of the gravitational field resembles that one obtained by Padmanabhan \cite{TP2} for a nongeodesic observer at rest in a static spacetime. That is not surprising because a static observer in a gravitational field is equivalent to an accelerated one in flat spacetime. \\
   Moreover, even the location of the radiation emitted by a uniformly accelerated charge may be explained using the nonstationarity of the Rindler metric beyond the horizon $X = 1/g$ (or $x_{M} = t_{M}$). It is a well established fact that an observer who is comoving with an accelerating charge will not detect any electromagnetic radiation since the radiation field is confined to a region inaccessible to him, namely $|x_{M}| < t_{M}$, where the geometry is time dependent (see \cite{AS, DB}). On the line of Edery - Constantineau view, we could assert that the classical radiation is localized in the nonstationary Milne spacetime, due to the continuous set of microstates parameterized by the time $t$. The radiation field (containing the radiating part extracted by separating the components dropping of as $1/r$ from the usual $1/r^{2}$ fields) have been obtained in many papers \cite{BG, FR, DB}. 
  
  To simplify the analysis, let us put $1 - gX = g \xi$ in the metric (2.1). One obtains
   \begin{equation}
 ds^{2} = - g^{2} \xi^{2} dT^{2} + d \xi^{2} + dY^{2} + dZ^{2}
 \label{4.6}
 \end{equation}
 We give here only the component $E^{\xi}$ of the static electric field in the (comoving with the charge) coordinates ($\xi$ is in the direction of acceleration). It is given by \cite{BG, FR}
 \begin{equation}
   E^{\xi} = \frac{4eg (\xi^{2} - \rho^{2} - g^{-2})}{[g^{2} (\xi^{2} + \rho^{2} + g^{-2})^{2} - 4\xi^{2}]^{3/2}} 
 \label{4.7}
 \end{equation}
 where $e$ is the charge of the electric particle and $\rho^{2} = Y^{2} + Z^{2}$.
 The same expression is obtained for the $x_{M}$ - component of the electric field in Minkowski coordinates, with only one difference: $\xi^{2}$ has to be replaced with $x_{M}^{2} - t_{M}^{2}$. We mention also that the lack of radiation in the comoving system is due to the null values of the components of the magnetic field which however are nonzero in the nonstatic region $\bar{x} > 1/2g$. On the surface $\xi = 0$ (the event horizon $H$) the electric field acquires the value
  \begin{equation}
   E^{\xi}_{H} = - \frac{4eg^{2}}{(1 + g^{2} \rho^{2})^{2}} 
 \label{4.8}
 \end{equation}
 with the surface charge density
   \begin{equation}
  \sigma_{H} = \frac{ E^{\xi}_{H}}{4 \pi} = - \frac{eg^{2}}{\pi (1 + g^{2} \rho^{2})^{2}} 
 \label{4.9}
 \end{equation}
 For small values of $Y, Z$ (or for $\rho << 1/g$), (4.9) yields 
    \begin{equation}
  \sigma_{H} \approx  - \frac{eg^{2}}{\pi}
 \label{4.10}
 \end{equation}
 We observe from (4.10) that $\pi/g^{2}$ plays the role of a surface (it is a portion of the horizon, where $|E^{\xi}_{H}|$ 
 acquires the highest values. Hence, taking $\pi/g^{2}$ as the area of the horizon, its entropy will be given by $S = area/4 = \pi/4g^{2}$, i.e. the value given Eq. (4.3), which has been obtained using different arguments.\\

 \section{Conclusions}
  The prescription of Edery and Constantineau is applied in this paper to prove that the Rindler horizon possesses microscopic DOF and, from here, an entropy proportional to $1/g^{2}$. It is rooted from the nonstationary character of the geometry beyond the event horizon which leads to a continuous set of classical microstates. The entropy measures the ignorance of a Rindler observer to know the value of the label $t \in (1/2g, \infty)$. 
  
  Using a congruence of  ''static'' geodesic observers labeled by the velocity field $u^{a}$, we have written the dynamic phase space $(h_{ab}, P^{ab})$ and the extrinsic curvature tensor of the hypersurface $\Sigma$ of constant time. The kinematical parameters of the congruence have a nontrivial form and $\dot{\Theta} < 0$, a necessary condition for the Raychaudhuri equation to be obeyed. Having a nonzero horizon energy, we computed it for our particular geometry and obtained a value already found in a previous paper.
  
  We further found the gravitational energy density is proportional to $g^{2}$. A similar dependence has been recently obtained by Padmanabhan for an observer at rest in a static field. In addition, the expression of the surface charge density $\sigma_{H}$ on the Rindler horizon gives us the possibility to get the same entropy formula as that obtained from the nonstationarity of the spacetime beyond the horizon.

\end{document}